# Long lifetime of thermally-excited magnons in bulk yttrium iron garnet

John S. Jamison[1], Zihao Yang[2], Brandon L. Giles[1], Jack T. Brangham[3], Guanzhong Wu[3], P. Chris Hammel[3], Fengyuan Yang[3], and Roberto C. Myers[1-3]

[1] *Department of Materials Science and Engineering,*

[2] *Department of Electrical and Computer Engineering,*

[3] *Department of Physics,*

*The Ohio State University, Columbus, OH, USA, 43210*

Spin currents are generated within the bulk of magnetic materials due to heat flow, an effect called intrinsic spin-Seebeck. This bulk bosonic spin current consists of a diffusing thermal magnon cloud, parametrized by the magnon chemical potential ($\mu_m$), with a diffusion length of several microns in yttrium iron garnet (YIG). Transient opto-thermal measurements of the spin-Seebeck effect (SSE) as a function of temperature reveal the time evolution of $\mu_m$ due to intrinsic SSE in YIG. The interface SSE develops at times < 2 ns while the intrinsic SSE signal continues to evolve at times > 500 µs, dominating the temperature dependence of SSE in bulk YIG. Time-dependent SSE data are fit to a multi-temperature model of coupled spin/heat transport using finite element method (FEM), where the magnon spin lifetime ($\tau$) and magnon-phonon thermalization time ($\tau_{mp}$) are used as fit parameters. From 300 K to 4 K, $\tau_{mp}$ varies from 1 to 10 ns, whereas $\tau$ varies from 2 to 60 µs with the spin lifetime peaking at 90 K. At low temperature, a reduction in $\tau$ is observed consistent with impurity relaxation reported in ferromagnetic resonance measurements. These results demonstrate that the thermal magnon cloud in YIG contains extremely low frequency magnons (~10 GHz) providing spectral insight to the microscopic scattering processes involved in magnon spin/heat diffusion.

## I. INTRODUCTION

The thermal generation of spin currents has stimulated a large body of theoretical and experimental work mainly focused on understanding the spin-Seebeck effect (SSE) [1–3]. An interface temperature difference between electrons in a normal metal and magnons in a magnetic insulator ($\Delta T_{me}$) drives a spin current. Heat flow within the bulk of the magnetic insulator also produces a spin current via the bulk or intrinsic SSE [4–8]. From steady-state measurements of these effects it is not possible to directly observe the dynamics of coupled magnon-electron spin transport, although meticulous magnetic field [9–11] and sample thickness [7,12,13] dependent measurements identified the energies and length scales of magnons contributing to SSE in yttrium iron garnet (YIG). These measurements agree that SSE magnons are low energy (< 1.5 meV). Furthermore, steady-state non-local spin transport measurements in YIG have isolated the spin diffusion length ($\lambda_m$) [8,14–16], revealing that the magnon chemical potential ($\mu_m$) decays on the micron length scale. Previously, SSE on the ultrafast time scale examined the interface magnon-electron temperature difference ($\Delta T_{me}$) [17–19]. However, low energy magnons involved in SSE evolve on much longer time scales. Microsecond timescale measurements by Agrawal *et al.* and Hioki *et al.* identified the contribution of bulk magnon transport [20–22], but did not examine the time evolution of $\mu_m$.

Here, we present time-dependent SSE measurements of the lifetime of non-equilibrium magnon spins $\tau$, the thermal magnon-phonon thermalization time $\tau_{mp}$, and the relative magnitude of the interface ($\Delta T_{me}$ driven) versus bulk ($\mu_m$ driven) components of the thermally-driven spin current. The data are fit to solutions of the coupled heat/spin transport equations [4,5] using a 2D axisymmetric finite element method (FEM). Intrinsic SSE exhibits a maximum at ~80 K, whereas interface SSE shows a comparatively weak temperature dependence. Surprisingly, $\tau$ varies from 2 to 60 μs, similar to the lifetimes of resonantly-excited (~ GHz) ferromagnetic resonance (FMR) modes. These results imply that the bulk spin currents within YIG contain a large population of extremely low frequency magnons. The results are consistent with the temporal evolution of the non-equilibrium Bose-Einstein (BE) distribution of thermal magnons in YIG. Namely, high-frequency (short lifetime) magnons rapidly cool from a distribution only described by the magnon temperature ($T_m$), to one described both by $T_m$ and $\mu_m$, which favors low-frequency (long-lived) magnons.

## II. TIME-RESOLVED OPTOTHERMAL SPIN-SEEBECK

### 1. Materials and Experimental Procedure

The sample is a polished single crystal of (100) YIG obtained from Princeton Scientific cut to 5×5×0.5 mm. The as–received YIG crystals are placed in an ultra-high-vacuum (UHV) off-axis sputtering chamber to deposit a buffer layer of YIG (30 nm), providing a clean new surface on which a Pt layer (6 nm) is deposited under the same UHV environment. This ensures the Pt/YIG interface is pristine. Optical lithography using a chlorine-based reactive ion etch is used to pattern a 50×50 μm square Pt active area with two 0.2 mm² contacts that are indium bonded to 25 −μm diameter annealed gold wires.

Utilizing an opto-thermal method [23], a 980 nm laser with a total rise time $\delta t = 3$ ns is digitally modulated at 1 kHz and focused onto the Pt film while a magnetic field $H_x$ is applied (Fig. 1(a)). The laser (~650 μW total absorbed power) is focused by a 40× objective to a spot ~6 μm in diameter onto the center of the active area. Spin currents are driven across the Pt/YIG interface by laser heating and detected as a transverse voltage drop across Pt due to the inverse spin Hall effect (ISHE) [24–26]. Fig. 1(b) plots the field-dependent transverse voltage $V_y$ measured using a lock-in amplifier referenced to the laser modulation frequency, where the SSE signal is defined as, $V_{SSE} = [V_y(+H) - V_y(-H)]/2$ [27]. SSE waveforms are acquired using an oscilloscope (KEYSIGHT DSOS204A) with a sampling rate of 1.25 GSa/s at an amplifier limited bandwidth of 350 MHz. The inset of Fig. 1(c) shows the time profile of the laser when it turns on at $t = 10.8$ ns (purple circles represent the 10-90% rise time of $2.2 \pm 0.2$ ns). Oscilloscope traces are taken at ±1.25 kOe noted by the blue and red squares in Fig. 1(b) and the blue and red curves in Fig. 1(c). At 30 K, $V_{SSE}$ resembles a square wave for which the root mean square (RMS) amplitude of the fundamental is $\sqrt{2}/\pi \sim 0.45$, matching the ratio between $V_{SSE}$ measured by lock-in amplifier ($V_{RMS}$) and oscilloscope: (3.8 µV)/(8.5 µV) ~ 0.45.

The measurement of $V_{SSE}$, shown in Fig. 1(c), is repeated at temperatures from 20-300 K. Figure 2(a) plots time-dependent $V_{SSE}$ at various temperatures. These data are normalized ($V_N(t) = V_{SSE}(t)/V_{SSE}(500 \text{ µs})$) and plotted on a logarithmic time-axis in Fig. 2(b). $V_N$ rises rapidly for ~3 ns independent of temperature (see inset Fig. 2(b)), while at later times $dV_N(t)/dt$ reduces, showing a gradual saturation out at $t > 100$ µs. The fast rise is associated with a rapidly generated $\Delta T_{me}$ [17]. The longer time scale is associated with the evolution of $\mu_m$. Due to the coupled spin/heat transport, it is not possible to isolate the time evolution of $\mu_m$ by simply subtracting the $\Delta T_{me}$ component.

## 2. Non-monotonic time-domain SSE

At low temperatures ($T \leq 30 \, K$) the $V_{SSE}$ signal exhibits a negative time derivative which persists over a time-scale of 100s of $\mu s$. At 4 K the total drop is 150 nV which is ~2% of the total signal (Fig. 3(a)), whereas at 10 K the drop is ~100 nV, ~1.5% of the total signal, (Fig. 3(b)). Beyond 10 K the non-monotonic behavior is observed by using is an adjacent averaging filter with a window size of 640 ns. Figs. 3(c) and (d) show the filtered $V_{SSE}$ waveform at 20 and 30 K, respectively. In Fig. 3(e) we plot the drop in $V_{SSE}(t)$ as a function of temperature. This filter is not used when fitting the data nor is it used in any other plots.

## III. COUPLED SPIN-HEAT TRANSPORT MODEL

### 1. Heat Transport

We use 2D axisymmetric FEM to first determine the temperature space/time profile within the sample. All thermal transport parameters are detailed in Appendix B1. The YIG is a 500×500 µm cylinder. The spin detector Pt is treated as a thin layer 30 μm in radius with a uniform temperature in the z direction (axial). The laser is treated as a volumetric heat source with a gaussian profile that is 3 μm in radius. A three-temperature (3T) model is implemented separating magnon temperature ($T_m$), phonon temperature ($T_p$), and electron temperature ($T_e$). Coupled magnon-phonon heat transport in the YIG is described by,

$$\begin{pmatrix} \rho C_m \frac{\partial T_m}{\partial t} - \kappa_m \nabla^2 T_m \\ \rho C_p \frac{\partial T_p}{\partial t} - \kappa_p \nabla^2 T_p \end{pmatrix} = \begin{pmatrix} G_{mp} & -G_{mp} \\ -G_{mp} & G_{mp} \end{pmatrix} \begin{pmatrix} T_m \\ T_p \end{pmatrix}, \qquad (1)$$

where $C_m$, $C_p$, $\kappa_m$, and $\kappa_p$ are the volumetric heat capacities and thermal conductivities for magnons and phonons in YIG, respectively, $\rho$ is the mass density of YIG, and $G_{mp}$ is the magnon-phonon coupling,

$$G_{mp} = \frac{\rho C_m C_p}{C_m + C_p} \left( \frac{1}{\tau_{mp}} \right). \qquad (2)$$

Electron-phonon coupling is assumed to be temperature independent at $10^{18}$ W/m³K [28]. Across the Pt/YIG interface, heat flows through magnon-electron conductance ($g_{me}$) and phonon-phonon conductance ($g_{pp}$) with a the total interface conductance $g_{th} = g_{me} + g_{pp}$, which we assume to be that of Au/Sapphire [29]. Continuity is maintained by $\kappa_e \partial_z T_e = \kappa_m \partial_z T_m$ and $\kappa_p^{Pt} \partial_z T_p^{Pt} = \kappa_p^{YIG} \partial_z T_p^{YIG}$. The interface and bulk thermal transport are related by

$$g_{me} = g_{th} \frac{\kappa_e \kappa_m}{\kappa_t^{YIG} \kappa_t^{Pt}}, \qquad (3)$$

where $\kappa_t^{YIG}$ and $\kappa_t^{Pt}$ are the total thermal conductivities of YIG and Pt, respectively.

## 2. Spin Transport

The thermal profiles are used as inputs to the spin diffusion equation in YIG, neglecting the spin-Peltier effect,

$$\frac{\partial \mu_m}{\partial t} = \frac{\lambda_m^2}{\tau}\left(\nabla^2 \mu_m + \frac{5\zeta\left(\frac{5}{2}\right)k_b}{2\zeta\left(\frac{3}{2}\right)^2}\nabla^2 T_m - \frac{\mu_m}{\lambda_m^2} - \frac{5\zeta\left(\frac{5}{2}\right)k_b}{2\zeta\left(\frac{3}{2}\right)\lambda_m^2}(T_m - T_p)\right), \tag{4}$$

where $\lambda_m$ was measured in this sample to be ~8 μm and temperature independent [8], and $\zeta(x)$ is the Riemann zeta function. In Eq. 4, $\nabla^2\mu_m$ describes spin diffusion, $\nabla^2 T_m$ describes spin generation via intrinsic SSE, $\mu_m/\lambda_m^2$ describes spin relaxation, and $k_b\Delta T_{mp}/\lambda_m^2$ describes the generation or relaxation of spin due to $\Delta T_{mp} = T_p - T_m$. Eq. 4 is derived using relationships between the magnon spin conductivity $\sigma_m$ and the intrinsic SSE coefficient $L/T$ in YIG [4,5] (Appendix A). The spin current density across the Pt/YIG interface is given by,

$$\boldsymbol{j}_s|_{int} = G\mu_m|_{int} + S\Delta T_{me}, \tag{5}$$

where $G$ is the interface spin conductivity and $S$ is the interface SSE coefficient. Theoretically, $S \approx Gk_b$, however, we find that at higher temperatures $S$ must be increased to capture the transient spin current on the short time scale, and therefore treat the $S/G$ ratio as an adjustable parameter. As the spin accumulation in the metal film evolves at $t \sim 10^{-12}$ s, [17] and our data span $t = 10^{-9}$ to $10^{-3}$ s, we neglect the electron spin chemical potential in Pt for computational efficiency. $G$ is determined by [4],

$$G = \frac{\hbar g^{\uparrow\downarrow}}{\pi\alpha s\tau}\frac{\partial \rho_m}{\partial \mu_m}, \tag{6}$$

where $g^{\uparrow\downarrow} = 1 \cdot 10^{18}$ $1/m^2$ is the spin mixing conductance, and $\alpha \approx 5 \times 10^{-5}$ is the Gilbert damping parameter for bulk YIG, measured by spin-hall magnetoresistance and FMR, respectively (Appendix B). There are numerous reports of a strong temperature dependence of the damping, [30–33] therefore the T-dependent $\alpha$ ought to be used in our modeling, however the literature is not consistent on the values, and indeed, the FMR based measurement of alpha is not valid below 100K [30]. Assuming a linear temperature dependence of $\alpha$ with a zero-temperature intercept, we carried out simulations of the time-domain SSE. However, in this case a $\mu_m$ reverses sign, and is in strong qualitative disagreement with the measurement (Appendix D). Therefore, we consider $\alpha$ to be temperature independent in the spin transport modeling. The saturation spin density is $s = 5$ nm$^{-3}$, $\rho_m$ is the non-equilibrium magnon density, and $\partial \rho_m / \partial \mu_m$ is evaluated at $\mu_m = 0$ by,

$$\frac{\partial \rho_m}{\partial \mu_m} = \sqrt{kT}\, Li_{\frac{1}{2}}\left(e^{-\frac{\Delta}{kT}}\right)(4\pi D_s)^{-\frac{3}{2}}, \tag{7}$$

where $\Delta = 14.8\ \mu eV$ is the magnon gap energy, and $D_s = 560\ meV \cdot Å^2$ is the spin stiffness [34,35].

## IV. RESULTS AND DISCUSSION

### 1. Thermal Transport Finite Element Modeling

In Fig. 4 we plot results of $T_m$ evaluated along the optic axis at the Pt/YIG interface.at 4, 90, and 300 K. The total magnon temperature excursion ($\Delta T_m$) behaves as expected; it saturates rapidly at low temperature and slowly at high temperature (Fig. 4(a)). $\partial T_m/\partial z$ (Fig. 4(b)), exhibits a different time dependence at 90 and 300 K, compared to 4 K. After the initial rise during the laser turn on time there is an additional rise over longer timescales. $\nabla^2 T_m$ (Fig. 4(c)) indicates that there is continuous magnon generation via intrinsic SSE near the interface across all time scales. At 300 K $\nabla^2 T_m$ becomes negative at $t = 150$ ns due to the non-equilibrium between magnons and phonons ($\Delta T_{mp} = T_p - T_m$). At low temperature, $\Delta T_{mp}$ peaks at 0.6 K. Interestingly, at high temperature the magnons are temporarily hotter than phonons $\Delta T_{mp} < 0$ which persists for approximately 20 ns. Even though the majority of the heat transport occurs through the phonon channel, the low $C_m$, compared to $C_p$, allows $T_m$ to temporarily exceed $T_p$.

## 2. Magnon Chemical Potential and Bulk Magnon Transport Via Intrinsic SSE

Fig. 5(a,b) plots the spatial distribution of $\mu_m$ at 90 K at 50 ns (a), and 500 μs (b). At 50 ns $\mu_m$ has only just begun to evolve at the interface. Below the building magnon depletion, a comparatively small magnon accumulation ($\mu_m > 0$) is generated by intrinsic SSE and spreads out spherically. At 500 μs, the magnon depletion has completely evolved and the magnon accumulation from earlier is absent, having spread out and decayed. Fig. 5(c) plots z-axis cuts of $\mu_m$ along the optical axis. Figure 6 plots the bulk spin currents within the YIG at the same times and location as Fig. 5(c). The intrinsic SSE current (Fig. 6(a)) points away from the interface, evolving with $\partial T_m / \partial T_z$. Interface SSE, driven by $S\Delta T_{me}$ is not as large as intrinsic SSE which results in a backflow spin current (and $\mu_m < 0$) towards the interface made up of non-equilibrium magnons as shown in Fig. 6(b). The total spin current within the bulk of the YIG generally flows away from the interface and peaks at a distance less than $\lambda_m$ from the interface.

Previously, we utilized a 1T model [36], however, this is inadequate in capturing the early rise time (1-10 ns) features of $V_N(t)$. A 3T model is critical to capture the SSE dynamics as magnons and phonons remain out-of-equilibrium over experimentally relevant timescales. Importantly, by including the magnon-phonon thermalization time ($\tau_{mp}$), non-monotonic features in the time-dependence of SSE are captured. For example, a ~2% drop in the SSE signal after the initial rise can be seen at 4 K (Fig. 3(a)). Since $T_p > T_m$ at the interface, then by Eq. (4) there is a generation of nonequilibrium magnons, reducing $|\mu_m|$ since $\mu_m < 0$. This interplay between magnon-phonon coupling and spin diffusion is observable in data up through 30 K. Removing the fourth term in Eq. (4) causes the negative time derivative in spin current to disappear entirely, indicating that $\Delta T_{mp}$ is parasitic to SSE at these conditions.

## 3. Fit Results

Using least squares regression, we fit the time-dependence of $j_s|_{int}(t)$ to $V_N(t)$ using $\tau$, $\tau_{mp}$, and $S/G$ as adjustable parameters. Figure 7 demonstrates the qualitatively distinct impact of the two timescales, $\tau$ and $\tau_{mp}$, on $\boldsymbol{j_s}|_{int}(t)$. At 4K, $\boldsymbol{j_s}|_{int}$ begins with a rapid rise of $S\Delta T_{me}$. Intrinsic SSE ($\nabla T_m$) at the interface is greater than $S\Delta T_{me}$, generating a local magnon depletion ($\mu_m < 0$), adding another term to $\boldsymbol{j_s}|_{int}$ ($G\mu_m|_{int}$). As $\tau$ decreases (increases), $\boldsymbol{j_s}|_{int}(t)$ shows a faster (slower) risetime.

The temperature dependence of $\tau$ (Fig. 8(a)) shows a peak of 65 ± 34 μs at 90 K decreasing to 15 ± 6 μs at 300 K. The increase in $\tau$ below 300 K correlates with the linear relationship between $\alpha$ and temperature. However, a direct connection between the FMR determined value of $\alpha$ and time-domain SSE measurement of $\tau$ is problematic. For example, at 300 K, with $\tau$ = 15 μs, assuming that $\tau$ represents the average lifetime of all the modes within $\rho_m$, then the average frequency, $f = (4\pi\alpha\tau)^{-1}$, would be 100 MHz, an unphysically low frequency, i.e. below the magnon gap, $f_\Delta$ = 3.5 GHz. This inconsistency is resolved by examining the spectral variation of magnon lifetimes [37], where magnon lifetimes increase strongly up to 25 μs at FMR frequencies at room temperature (Fig. 9(a)). This lifetime variation can be used to predict the spectral variation of magnon spin conductance, $\sigma_m(\omega) = \hbar\lambda_m^2\partial\rho(\omega)/\partial\mu(\tau(\omega))^{-1}$ (Fig. 9(b)). Taking the integral of this function over the principle magnon band (Fig. 8(c)), it is observed that roughly half of the spin conductance arises from magnons with frequencies of 100 GHZ or less. Although the low frequency magnons have far higher lifetimes, their low mobility limits their conductance, thus in YIG, both high and low frequency magnons are involved in spin transport at room temperature.

It is now well established that FMR magnons are produced during spin-heat transport in YIG. Indeed, as magnons are bosons and obey BE-statistics, they can be condensed into a BEC phase [4,38,39] These recent experimental and theoretical works demonstrate the thermal excitation of magnons and condensation into low lying FMR frequencies and clearly establish the magnon chemical potential paradigm that spans the entire frequency range [4,5]. Furthermore, magnetic field dependent measurements of SSE and non-local spin transport indicate that many of the magnons contributing to SSE are sub-thermal (~1 meV) [9–11]. Adding to this body of evidence we observe a decrease of $\tau$ below 90 K, reaching 3 ± 0.3 µs at 4 K (Fig. 9(a)). This is similar to the increase in FMR linewidths observed over the same temperatures for low frequency magnons (10-40 GHz), an effect previously attributed to impurity relaxation [30,31,40]. The idea that these extremely low energy magnons contribute to most of the transport does not explain the observed temperature dependence of Gadolinium Iron Garnet (GdIG) [41], however the experiments performed by Geprägs *et al.* are performed primarily on nm-thick films and the maximum thickness is 1 µm. The low frequency magnons originate from long distance transport in the bulk and only become significant when the sample thickness is greater than $\lambda_m$. To our knowledge no systematic examination of $\lambda_m$ has been performed in GdIG, however if it is comparable to YIG even within order of magnitude we would not expect any significant contribution from low energy magnons on this scale.

Based on the measured $\lambda_m$ [8,15] and $\tau$ (Fig. 9(a)), we calculate the magnon spin conductivity $\sigma_m$ using the Einstein relation,

$$\sigma_m = \hbar \frac{\lambda_m^2}{\tau} \frac{\partial \rho_m}{\partial \mu_m}. \qquad (8)$$

It is observed to increase with temperature above 100 K (Fig. 8(c)). Although our values of $\sigma_m$ are determined from time-domain SSE, they are in remarkable agreement with those determined by non-local steady-state measurements [15]. The agreement between steady-state non-local spin transport FEM modeling and time-domain SSE modeling provides strong validation of FEM modeling of coupled spin/heat transport. We note that the experimental geometries (boundary conditions) and measurement modalities are quite distinct. We also include the room temperature predicted total spin conductance based on Rückriegel et al. by integrating the spectral spin conductance (Fig. 9(c)) which agrees with both our data and that of Cornelissen *et al.* at room temperature.

The parasitic contribution to SSE is gauged by $\tau_{mp}$ (Fig. 9(b)). For T > 20 K, $\tau_{mp}$ is ~10 ns. Above 20 K, $\tau_{mp}$ slows the rate at which $\Delta T_{me}$ evolves. This is the main parameter affecting the SSE waveform for 1 ns < t < 100 ns. The interface $T_p$ continues evolving over longer times; through magnon-phonon coupling the evolution of $T_p$ contributes to the evolution of $\partial T_m/\partial z$ and therefore of $\Delta T_{me}$ as well. The long $\tau_{mp}$ values (~10 ns) are surprising as the magnon-phonon scattering time is estimated to be ~1-10 ps for thermal magnons [42]. However, a second length scale for SSE in YIG, the magnon-phonon thermalization length $\lambda_{mp}$,

$$\lambda_{mp} = \left[G_{mp}\left(\kappa_m^{-1} + \kappa_p^{-1}\right)\right]^{-\frac{1}{2}}, \quad (9)$$

was reported to be 100 nm at elevated temperatures [7]. Based on our results, we estimate $\lambda_{mp}$ = 300 nm at room temperature, in close agreement with Prakash *et al.*, supporting the long values of $\tau_{mp}$.

Figure 9(d) plots the time dependence of the two spin current channels at 300 K. Except for the non-monotonic behavior at low temperature, the general time domain shape of $S\Delta T_{me}$ and $G\mu_m$ is uniform across all temperatures. $S\Delta T_{me}$ dominates the spin current for t < 200 ns, and $G\mu_m$ always dominates the time domain signal for t > 1 μs. The contribution of both $S\Delta T_{me}$ and $G\mu_m$ to the steady-state SSE signal is determined by multiplying the normalized spin currents by $V_{SSE}$ at 500 μs. Below 40 K, $V_{S\Delta T}$ is the primary contribution to SSE (Fig. 9(d)). At intermediate temperatures, where $V_{SSE}$ is maximal, $V_{G\mu}$ dominates, and around 150 K the situation reverses again. Overall, the primary driver of the temperature dependence of $V_{SSE}$ is $\mu_m$.

## V. CONCLUSIONS

In summary, $\mu_m$ evolves over > 500 μs in bulk YIG compared with interface SSE that evolves at times <100 ns. The lifetime of magnons due to SSE peaks at 60 $\mu s$ at 90 K. Its temperature trend is similar to that of the FMR linewidth in YIG [34]. This connection between FMR (~10 GHz) and SSE (10 GHz to 12 THz) is understood as due to the spectral distribution of magnons skewed selectively toward very low frequencies in YIG. The spectral variation of spin conductance in YIG is understood in terms of the balance between hot THz magnons (high mobility / short lifetimes) and cold GHz magnons (low mobility / long lifetimes), such that intermediate range 10-50 GHz (exchange-dipolar magnons) provide a large contribution to bulk spin conductance in YIG. From the Einstein relation for magnons, the long lifetimes are in quantitative agreement with $\sigma_m$ observed in steady-state non-local measurements. This agreement between steady-state spin transport and time-domain SSE provides independent validation of the FEM modeling of coupled spin-heat transport. It is also observed that the non-equilibrium between magnons and phonons is observed to be parasitic to SSE. Finally, we find that in bulk YIG, the largest source of spin current crossing the Pt/YIG interface is $\mu_m$, which also dominates the temperature dependence of SSE.

**Acknowledgments:**

We thank Joseph Heremans, L.J. Cornelissen, R.A. Duine, David G. Cahill, Joseph Barker, and Andreas Rückriegel for providing valuable insight during discussions. This work was supported primarily by the Center for Emergent Materials at The Ohio State University, an NSF MRSEC (Award Number DMR-1420451) and by the Army Research Office through MURI W911NF-14-1-0016. Partial support is provided by Department of Energy under Grant No. DE-SC0001304 (JTB) and DE-FG02–03ER46054 (GW).

**APPENDIX A: COUPLED SPIN-HEAT TRANSPORT EQUATIONS**

The magnon spin current in the YIG $\boldsymbol{j}_s$, in units of $J/m^2$, can be expressed as

$$\boldsymbol{j}_s = -\left(\sigma_m \nabla \boldsymbol{\mu}_m + \frac{L}{T}\nabla \boldsymbol{T}_m\right), \tag{A1}$$

where $\sigma_m$ is the spin conductivity in units of $1/m$ and $L/T$ is the bulk spin Seebeck (SSE) coefficient in units of $Jm^{-1}K^{-1}$, and $\mu_m$ the magnon chemical potential in $J$. [5]. The continuity relation reads

$$\frac{\partial \rho_m}{\partial t} + \frac{1}{\hbar}\nabla \cdot \boldsymbol{j}_s = -\frac{\mu_m}{\tau}\frac{\partial \rho_m}{\partial \mu_m} + \Gamma_{\rho T}C_m \Delta T_{mp}, \tag{A2}$$

where $\rho_m$ is the non-equilibrium magnon density, $\tau$ is the lifetime of the magnon chemical potential, $\Gamma_{\rho T}$ is the relaxation rate of the non-equilibrium density by the magnon phonon temperature difference $\Delta T_{mp} = T_p - T_m$. By combining Eqs. (A1) and (A2) and assuming that variations in $\sigma_m$ and $L/T$ are negligible we get

$$\frac{\partial \rho_m}{\partial t} = \left(\sigma \nabla^2 \mu_m + \frac{L}{T}\nabla^2 T_m\right) - \frac{\mu_m}{\tau}\frac{\partial \rho_m}{\partial \mu_m} + \Gamma_{\rho T}C_m \Delta T_{mp}. \tag{A3}$$

The length scale over which the magnon chemical potential decays in steady state is also known as the spin diffusion length $\lambda_m$ and is related to the magnons spin conductivity as

$$\sigma_m = \frac{\lambda_m^2}{\tau}\frac{\partial \rho_m}{\partial \mu_m}. \tag{A4}$$

To reduce the continuity equation further, a relationship between $\sigma_m$ and $L/T$ is utilized from the theory described in Ref. [5].

$$\sigma_m = 4\frac{\zeta\left(\frac{3}{2}\right)^2 D_s \tau_s}{\hbar^2 \Lambda^3} \tag{A5}$$

$$\frac{L}{T} = 10\frac{\zeta\left(\frac{5}{2}\right) k_b D_s \tau_s}{\hbar^2 \Lambda^3} \tag{A6}$$

The spin conductivity and bulk SSE coefficients are described in terms of the YIG spin stiffness $D_s$, the mean scattering time $\tau_s$, where magnon phonon scattering is assumed to be the dominant mechanism, and the magnon thermal de Broglie wavelength $\Lambda = \sqrt{4\pi D_s/kT_m}$. By substituting Eq. (A5) into (A6) we get

$$\frac{L}{T} = \sigma_m \frac{5\zeta\left(\frac{5}{2}\right) k_b}{2\zeta\left(\frac{3}{2}\right)^2} . \tag{A7}$$

Additional reduction of the continuity relation is achieved for the fourth term on the right-hand side of Eq. (S3) by relating $\Gamma_{\rho T}$ to $\lambda_m$ and $\sigma_m$,

$$\Gamma_{\rho T} = \frac{5\zeta\left(\frac{5}{2}\right) k_b \sigma_m}{2\zeta\left(\frac{3}{2}\right) \lambda_m^2 C_m} \tag{A8}$$

Finally, by substituting Eqs. (A4), (A7), and (A8) into Eq. (A3) and dividing through by $\partial \rho_m/\partial \mu_m$ we arrive at

$$\frac{\partial \mu_m}{\partial t_m} = \frac{\lambda_m^2}{\tau_m}\left(\nabla^2 \mu_m + \frac{5\zeta\left(\frac{5}{2}\right) k_b}{2\zeta\left(\frac{3}{2}\right)^2}\nabla^2 T_m - \frac{\mu_m}{\lambda_m^2} + \frac{5\zeta\left(\frac{5}{2}\right) k_b}{2\zeta\left(\frac{3}{2}\right)\lambda_m^2}\Delta T_{mp}\right), \tag{A9}$$

which is Eq. (3) in the main text. This allows us to use three free parameters, $\tau$, $\tau_{mp}$, and the $S/G$ ratio, to fit the time profile of the measured spin current crossing the interface.

**APPENDIX B: TRANSPORT PARAMETERS**

**1. Thermal Transport**

Eq. (1) in the main text describes the heat transport in YIG where $T_m$ is coupled to $T_p$ via $G_{mp}$. Below ~20 K the magnon heat capacity $C_m$, and magnon thermal conductivity $\kappa_m$ are taken from Ref. [43]. Beyond 20 K these parameters are not measured and in the case of $\kappa_m$ the values are logarithmically extrapolated from Ref. [43]. The values used in simulations are plotted in Fig. 10(a). For $C_m$, plotted in Fig. 10(c), we utilized the $T_m^{\frac{3}{2}}$ law where $C_m$ follows the following relation,

$$C_m = \frac{15\zeta\left(\frac{5}{2}\right)}{32\,\rho_{YIG}}\sqrt{\frac{k_b^5 T_m^3}{\pi^3 D_s^3}}, \tag{A10}$$

where $\rho_{YIG} = 5170\ kg/m^3$ is the mass density of YIG at room temperature and $D_s = 560\ meV \cdot Å^2$ is the spin stiffness [35,44,45]. Within the FEM we consider the temperature variation in $\rho_{YIG}$ by linearly interpolating the measured lattice constants between 4 K and 300 K.

The total thermal conductivity for thin film Pt is taken from Ref. [46] and logarithmically extrapolated for temperatures below ~20 K. The electron thermal conductivity $\kappa_e$ is separated from the phonon thermal conductivity $\kappa_p^{Pt}$ in Pt by making use of the Wiedemann-Franz law,

$$\kappa_e = L_0 \sigma_e T, \tag{A11}$$

where $\sigma_e$ is the electrical conductivity in Pt taken from Ref. [46] in the same manner as the total thermal conductivity, and $L_0 = 2.44 \times 10^{-8}\ W\Omega\ K^{-2}$ is the Sommerfeld value for the Lorenz number. The total thermal conductivity in Pt is then treated as the sum of $\kappa_e$ and $\kappa_p^{Pt}$, and their values are plotted in Figure 10(b). $C_e$ is calculated from the linear relationship,

$$C_e = \gamma_e T, \tag{A12}$$

where $\gamma_e = 0.034\ JK^{-2}kg^{-1}$ [47]. The total heat capacity of Pt [47] is taken to be the sum of $C_e$ and $C_p^{Pt}$ plotted in Fig. 10(d).

In the absence of data of the interface thermal conductance for Pt/YIG, we use that of Au/Sapphire ($g_{th}$= 50 MWm$^{-2}$K$^{-1}$) from Ref. [29]. Since it varies weakly with temperature, we consider it to be independent of temperature. This only affects the absolute value of temperature rises, $\Delta T_{me}$, and $\Delta T_{pp}$, but not any of the gradients of temperature or other quantities critical to the modeling of the spin transport.

## 2. Experimental Determination of Gilbert Damping and Spin-Mixing Conductance

Within the FEM model, the spin current which crosses the interface is proportional to $\Delta \mu \approx \mu_m$, and $\Delta T_{me}$, where the constants of proportionality are $G$ and $S$, respectively. From Ref. [4], we can estimate $G$ to be

$$G = \frac{\hbar g^{\uparrow\downarrow}}{\pi \alpha s \tau} \frac{\partial \rho_m}{\partial \mu_m}, \tag{A13}$$

where $\alpha$ is the Gilbert damping parameter and $g^{\uparrow\downarrow}$ is the spin mixing conductance. To more accurately estimate $G$ we perform ferromagnetic resonance (FMR) and spin-Hall magnetoresistance (SMR) measurements on the real sample used in the experiments. For the FMR experiments the YIG crystal was placed over a microwave stripline, the field pointing out of plane, and $dI/dH$ was taken for driving frequencies ranging from 9 to 12 $GHz$. The microwave power used was $-10\ dBm$. Numerous resonances were detected and many shift past one another making determining $\Delta H$ difficult across multiple frequencies. However this is somewhat simplified since $\alpha$ should be the same for all modes, whereas only the inhomogeneous broadening varies [48]. Figure 11(a-c) shows the absorption data for 9, 11, and 12 $GHz$, respectively.

The Gilbert damping can be calculated by fitting the resonance spectra to,

$$\Delta H = \Delta H_0 + \frac{4\pi \alpha f}{\gamma}, \tag{A14}$$

where $f$ is the resonance frequency, $\Delta H$ is the measured linewidth of absorption, and $\gamma \approx 28\ GHz/T$ is the gyromagnetic ratio [30,49]. Figure 12(a,b) shows the resonance field and linewidths vs frequency, respectively. By fitting the data to Eq. (A14) we find $\alpha = 5 \cdot 10^{-5}$ which is consistent with high quality YIG.

Spin-Hall Magnetoresistance measurements were performed on the same Pt/YIG sample as used in the experiments described in the main text. A four-wire measurement of the resistivity of the Pt film is performed as a function of applied field $H_x$. In this case the applied field is also along the direction of the measured resistivity $\rho$. The measured change in resistivity $\Delta\rho_1$, shown in Fig. 13, is related to $g^{\uparrow\downarrow}$ by

$$\frac{\Delta\rho_1}{\rho} = \theta_{SH}^2 \frac{\lambda_{Pt}}{d} \frac{2\lambda_{Pt} g^{\uparrow\downarrow} \tanh^2 \frac{d}{2\lambda_{Pt}}}{\sigma + 2\lambda_{Pt} g^{\uparrow\downarrow} \coth \frac{d}{\lambda_{Pt}}}, \tag{A15a}$$

$$\rho = \rho_0 + \Delta\rho_0 + \Delta\rho_1(1 - m_y^2), \tag{A15b}$$

where $\theta_{SH} = 0.1$ is the spin-Hall angle of Pt, $\lambda_{Pt} = 7\ nm$ is the spin diffusion length of Pt [50,51], and $d = 6\ nm$ is the thickness of the Pt layer. $\Delta\rho_1/\rho \approx 4 \cdot 10^{-4}$ which results in $g^{\uparrow\downarrow} \approx 1 \cdot 10^{18}$.

determined by the Hessian matrix for SSR. The diagonal elements of the inverted Hessian are the standard variances for each fitting parameter.

### APPENDIX C: NON-EQUILIBRIUM MAGNON SPECTRUM

The spectrally resolved non-equilibrium magnon density $\rho_m'$ can be analytically evaluated in a parabolic single band model by

$$\rho_m'(\omega, \mu) = \frac{1}{4\pi^2} \left(\frac{\hbar}{D_s}\right)^{\frac{3}{2}} \sqrt{\omega} [\langle n_k(\omega, \mu_m, T_m)\rangle - \langle n_k(\omega, 0, T_m)\rangle], \tag{A16}$$

where $\langle n_k(\omega, \mu, T_m)\rangle$ is the Bose-Einstein distribution evaluated at some frequency $\omega$ and chemical potential.

In Fig. 14(a,b) we plot $\rho_m'$ at $T_m = 20$ K and $\mu_m = +1$ and $-1$ µeV, respectively. The magnon gap frequency, 3.5 GHz, is determined by the applied field in the experiment. The spectra in Fig. 14(a,b) may be numerically integrated and differentiated with respect to $\mu_m$ to give $\partial \rho_m / \partial \mu_m$. The analytical solution to $\partial \rho_m / \partial \mu_m$ at $\mu_m = 0$ is given by Eq. (6) in the main text. Regardless of the value of $\mu_m$, $\partial \rho_m / \partial \mu_m$ is linear with temperature so long as $k_b T \gg |\mu_m|$, Fig. 14(c). The peak ($\omega_p$) and mean ($\omega_m$) frequencies can be evaluated numerically from the spectra generated by Eq. (A16). Interestingly, at a given value of $\mu_m$, $\omega_p$ exhibits almost zero dependence on temperature, Fig. 15(a,b). Even under very large negative $\mu_m$, $\omega_p$ remains within a factor of 2 of the magnon gap energy. $\omega_m$ however, varies strongly with temperature as well as with the value of $\mu_m$, Fig. 15(c,d). $\omega_m$ in these calculations is generally around 100-200 GHz, although at temperatures greater than 20 K the $\omega_m$ would be underestimated in a parabolic band model.

## APPENDIX D: VARIATION OF SPIN TRANSPORT WITH INPUT PARAMETERS

### 1. Low Temperature Scaling of $\alpha$

Within the main text we discuss the effect which large $G$, relative to $\sigma_m/\lambda_m$ [4], has on the time domain spin transport at low temperature. Theoretically, $\alpha$ scales linearly with temperature. From Eq. (6), $G$ and $S$ scale linearly with the ratio $g^{\uparrow\downarrow}/\alpha$, therefore, at 4 K, $g^{\uparrow\downarrow}/\alpha$ is 75 times larger than at 300 K. However, $\partial \rho_m / \partial \mu_m$ also scales with temperature, with the ratio of $\partial \rho_m / \partial \mu_m (300\ K) / \partial \rho_m / \partial \mu_m\ (4\ K)$ being 87.5. This still results in lower interfacial spin transport coefficients at low temperature, as expected theoretically. But within this regime the interfacial spin transport coefficients become a much stronger influence on the time evolution of SSE compared to the bulk transport coefficients and we lose sensitivity to both $\tau$ and $\tau_{mp}$. Furthermore, the qualitative time dependence of the spin transport completely changes. Assuming a temperature dependent $\alpha$, Fig. 16 plots the modeled SSE waveform compared with the measured 4 K $V_N$. The long time-scale evolution of SSE in this case has the complete opposite time derivative over the entire time range, except for the initial rise. Additionally, the sign of $\mu_m$ within the entire structure changes from negative to positive due to spin injection from the Pt. The use of a T-dependent linear scaled $\alpha$ is therefore unsuitable in the low temperature regime. Making matters worse, recent experimental work indicates that FMR based measurements of Gilbert damping below 100 K are untenable [34].

## 2. High Temperature Thermal Magnon Transport Coefficients

Recently, predictions of $C_m$ have been made implementing quantum thermal statistics to the atomistic spin dynamics in YIG [52]. The calculations include all magnon bands and are calculated over the entire Brillouin Zone. Unsurprisingly, the predicted $C_m$ is much higher than that which is predicted by a $T^{\frac{3}{2}}$ law, 14.2 and 3 J/kgK, respectively. Here we investigate the effect on the time domain transport, and the fits at this new $C_m$. In Fig. 17 we show the best fit at 300 K under this $C_m$. Here $\tau$= 1 ns, $\tau_{mp}$=250 $\mu$s and $S/G$=9. In this case the $\tau_{mp}$ would be unreasonable, and $\lambda_{mp}$ would be 15 $\mu$m. Agrawal *et al.* [53] used BLS measurements to optically probe the length scale of $\lambda_{mp}$. Were $\lambda_{mp}$ on such a length scale it would be visible under BLS measurements but no such length scale in $\Delta T_{mp}$ was observed. Additionally the magnitude of $\lambda_{mp}$ is in major disagreement with the results from Prakash *et al.* [7]. As stated previously, the relationships between the various spin and heat transport parameters used in the FEM simulations are taken from single parabolic band approximations, whereas the calculations from Barker *et al.* utilize a fully quantum model which include all the bands. When the heat capacity is increased to include all of these bands there may be artifacts introduced in the fitting due to the interplay between parameters which take into account only single parabolic bands and parameters which include multiple bands. In particular, this could be from the fact that such heat capacities would naturally include the heat capacity from the anti-ferromagnetic modes which would be parasitic to the net thermal spin transport, but they are not included in the models used here to calculate our spin currents. More work is needed to construct spin-heat diffusion models which take into account multiple magnon bands.

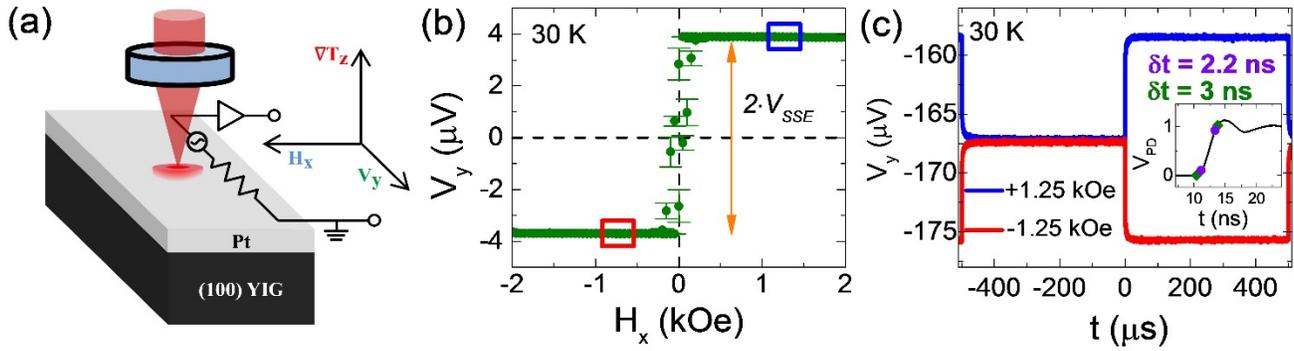

FIG. 1. (Color Online) (a) Schematic of the experimental setup. (b) $V_y$ measured vs. magnetic using a lock-in amplifier at 30K. (c) $V_y$ measured as a function of time and magnetic field using the oscilloscope at 30 K. Inset: Photodiode response to laser turning on at t = 10.8 ns and it's 10-90% rise time to be 2.2 ns and its 100% rise time to be 3 ns.

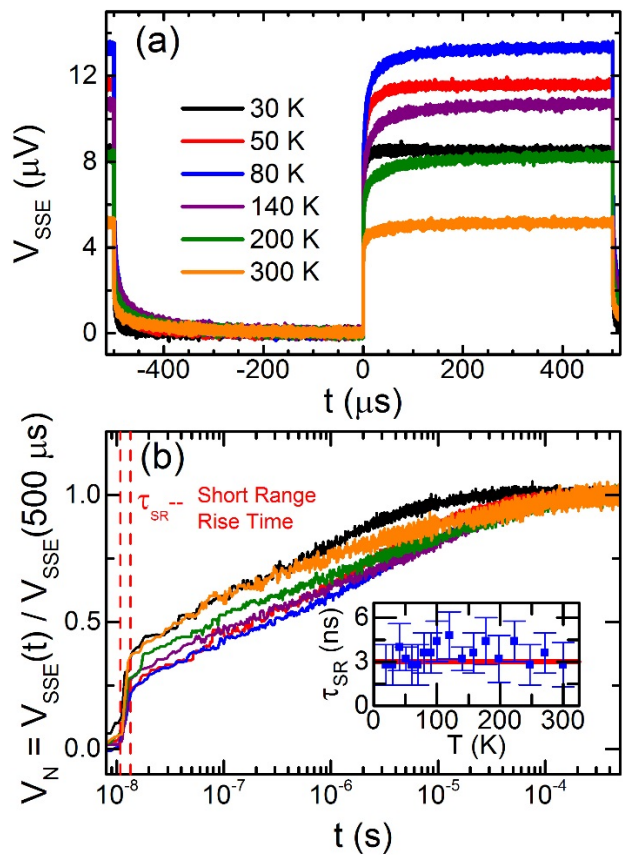

FIG. 2. (Color Online) (a) $V_{SSE}$ waveform vs time at various temperatures. (b) same data set as (a) normalized by $V_{SSE}(500\mu s)$ and plotted on a logarithmic time axis. Inset: $\tau_{SR}$ as a function of temperature the red horizontal line represents the 0-100% rise time of the laser.

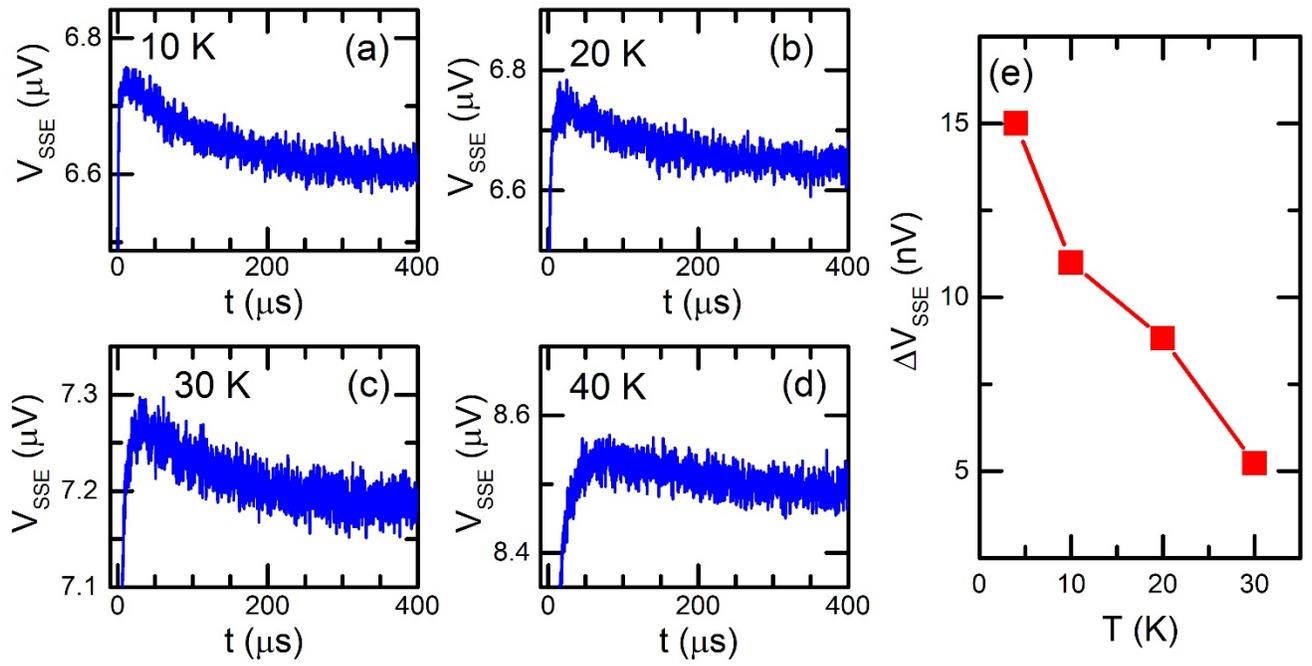

FIG. 3. (Color Online) Zoom of $V_{SSE}$ versus time to show the negative $\partial V_{SSE}/\partial t$ exhibited over several hundred microseconds at 4 (a), 10 (b), 20 (c), and 30(d) K. The data in (a-d) are filtered using an adjacent averaging filter with a window width of 640 ns. (e) Drop in SSE signal ($\Delta V_{SSE}$) as a function of temperature as defined in the figure.

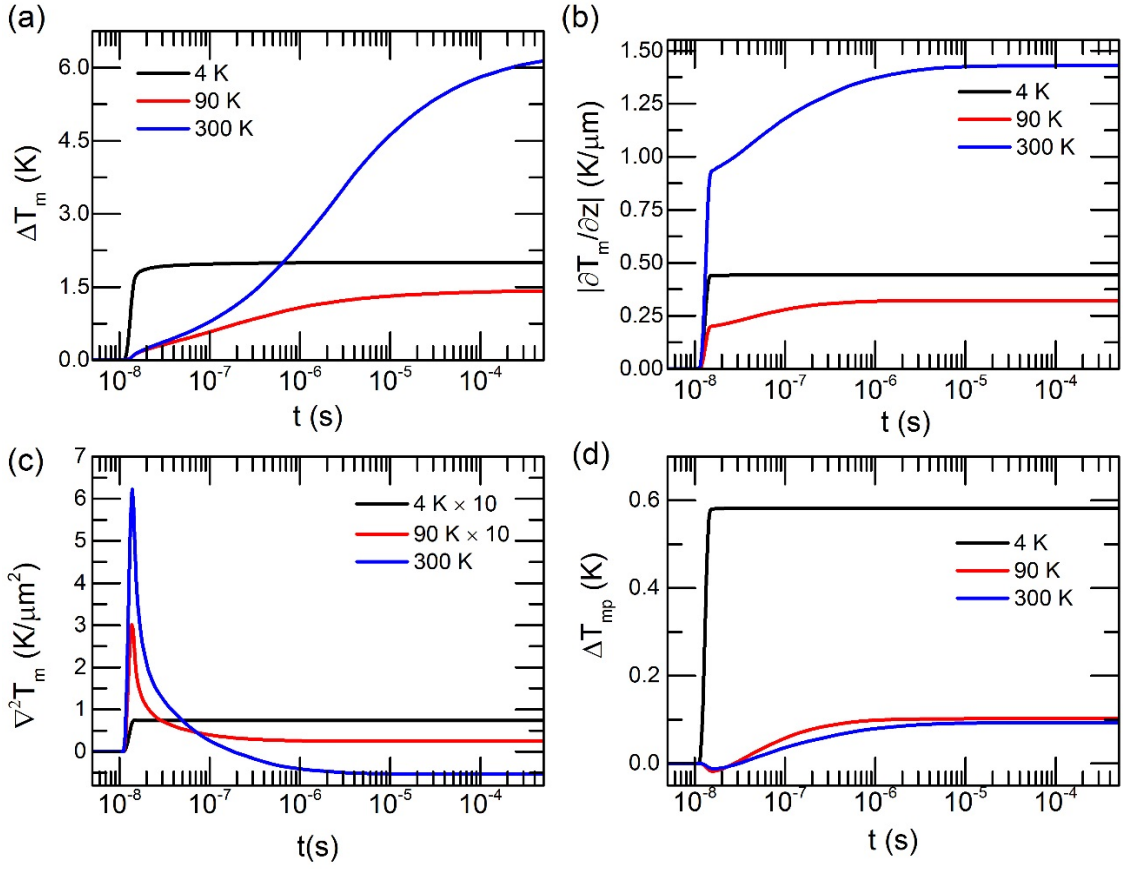

FIG. 4. (Color Online) Total magnon temperature excursion $\Delta T_m$ (a), z-component of the magnon temperature gradient (b), Laplacian of magnon temperature (c), and magnon phonon temperature difference (d) at 4 (black), 90 (red) and 300 K (blue). Data are taken from the FEM simulations at $r = 0$ and at $z = 0$.

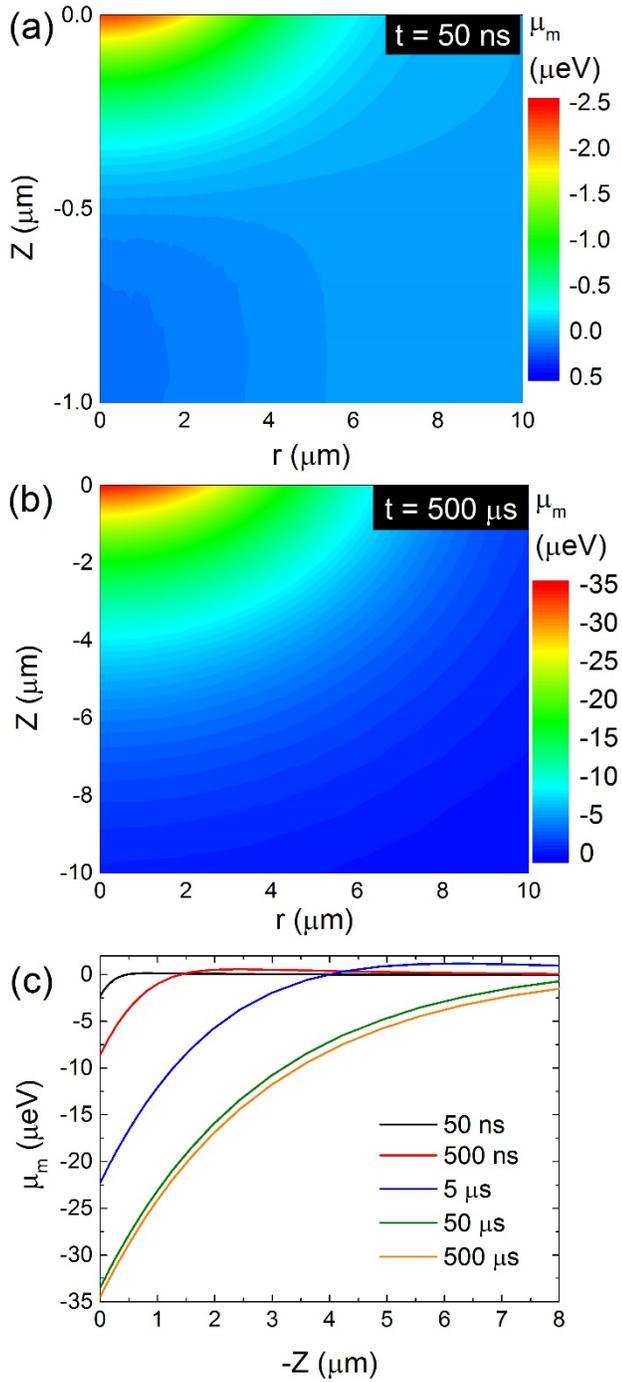

FIG. 5. (Color Online) Contour maps of $\mu_m$ in YIG at $t = 50$ ns (a), and $t = 500$ μs (b) taken from the FEM simulations at the best fit. Note the different color scales and y axis ranges between (a) and (b). (c) z axis profiles of $\mu_m$ taken along the optical axis from the FEM simulations at various times.

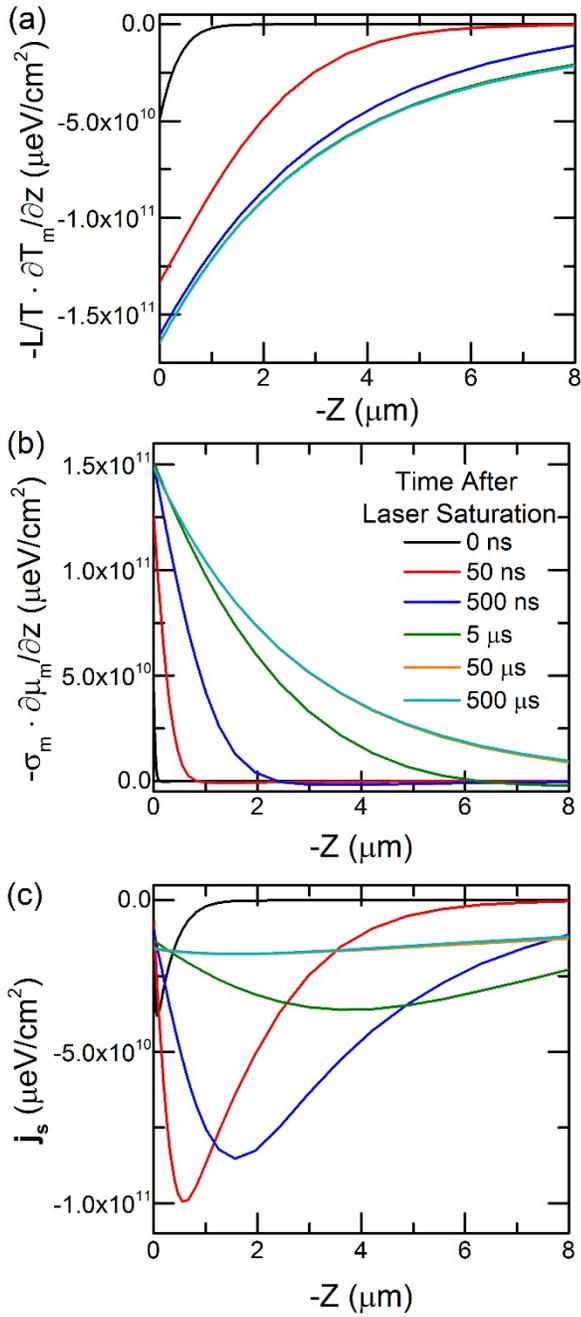

FIG. 6. (Color Online) z-axis cuts of the bulk intrinsic SSE spin current (a), bulk non-equilibrium magnon spin current (b), and total bulk spin current (c) in YIG taken from the FEM simulations at the best fit at 90 K. (a-c) share a legend.

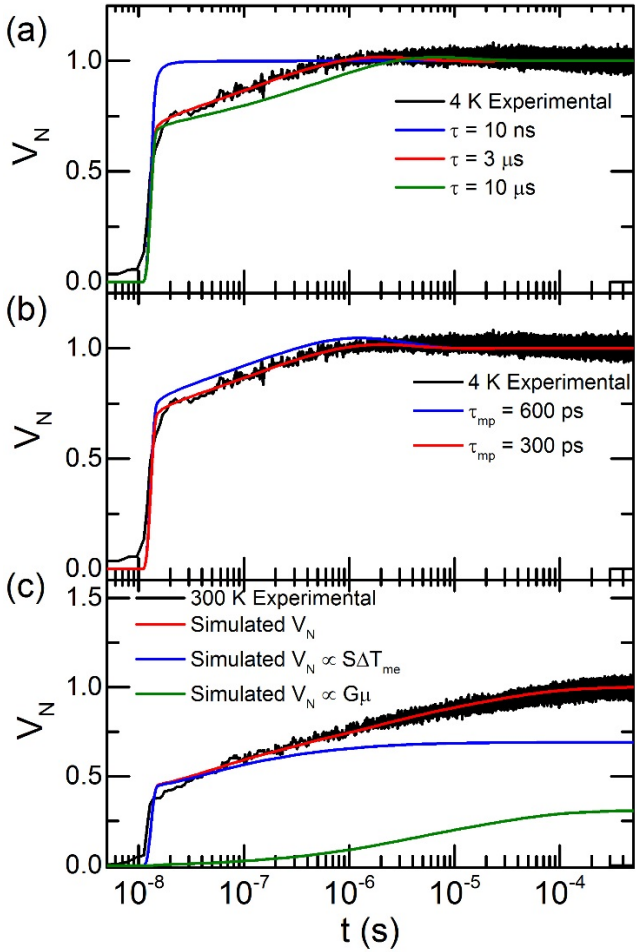

FIG. 7. (Color Online) Measured $V_N$ (black) at (a-b) 4 K and (c) 300 K. Modeled $V_N$ are plotted for different values of (a) magnon spin lifetime $\tau$ and (b) magnon-phonon thermalization time $\tau_{mp}$, in order to show the impact on the SSE waveform. (c) The simulated $V_N$ (red) is separated into the bulk $G\mu$ (green) and interface $S\Delta T_{me}$ (green) components.

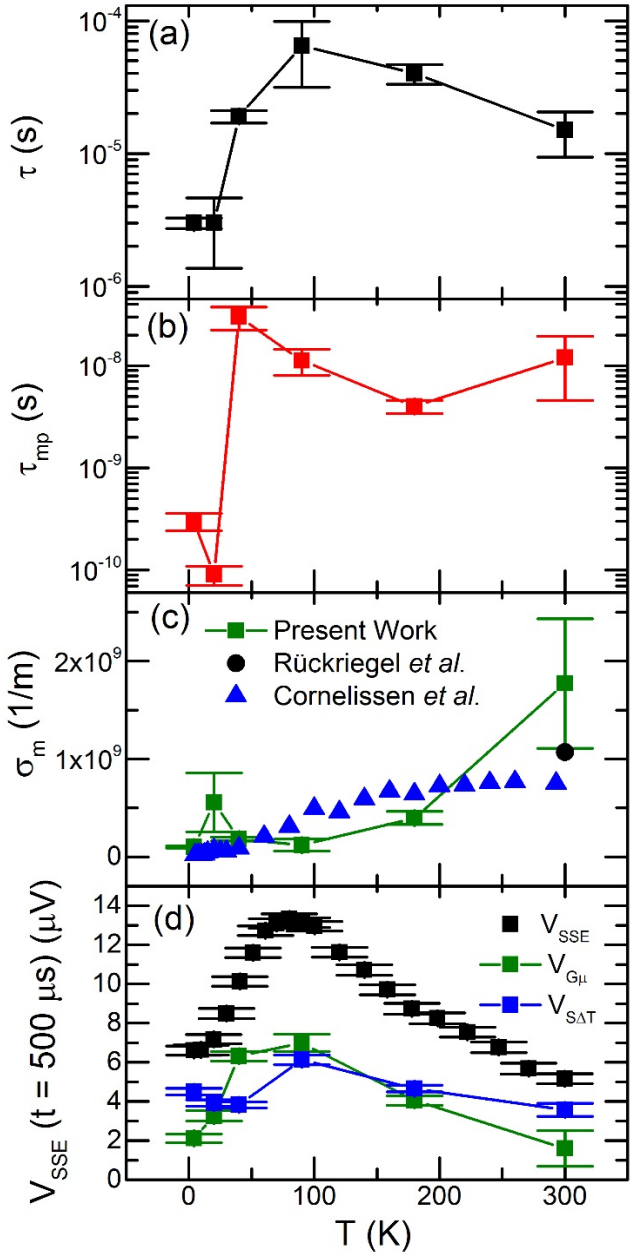

FIG. 8. (Color Online) Temperature dependences of (a) magnon spin lifetime $\tau$, (b), magnon-phonon thermalization time $\tau_{mp}$, (c) magnon spin conductivity $\sigma_m$, and (d) pseudo-steady-state SSE voltage: measured $V_{SSE}$ ($t = 500$ μs) (black), modeled SSE due to $\mu_m$ (green), and $\Delta T_{me}$ (blue). Error bars are one standard error of the fits in each panel.

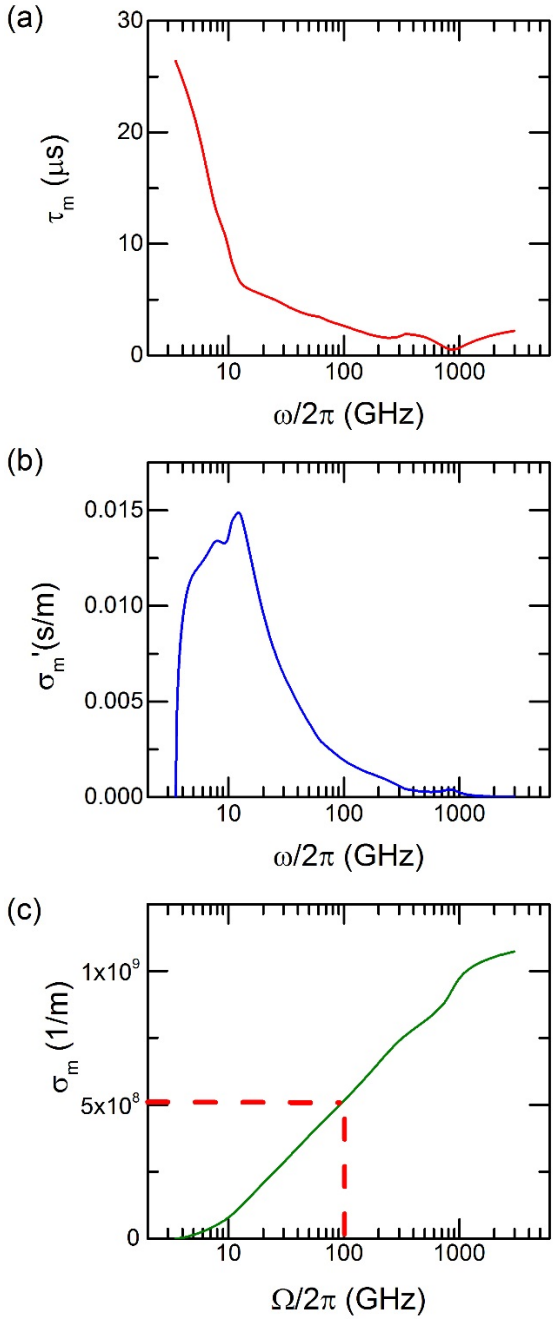

FIG. 9. (Color Online) (a) Spectrally resolved magnon relaxation times from Ref. [37]. (b) Spectrally resolved magnon spin conductivity calculated from a frequency dependent Eq. (8) and (a). (c) Cumulative magnon spin conductivity calculated from (b).

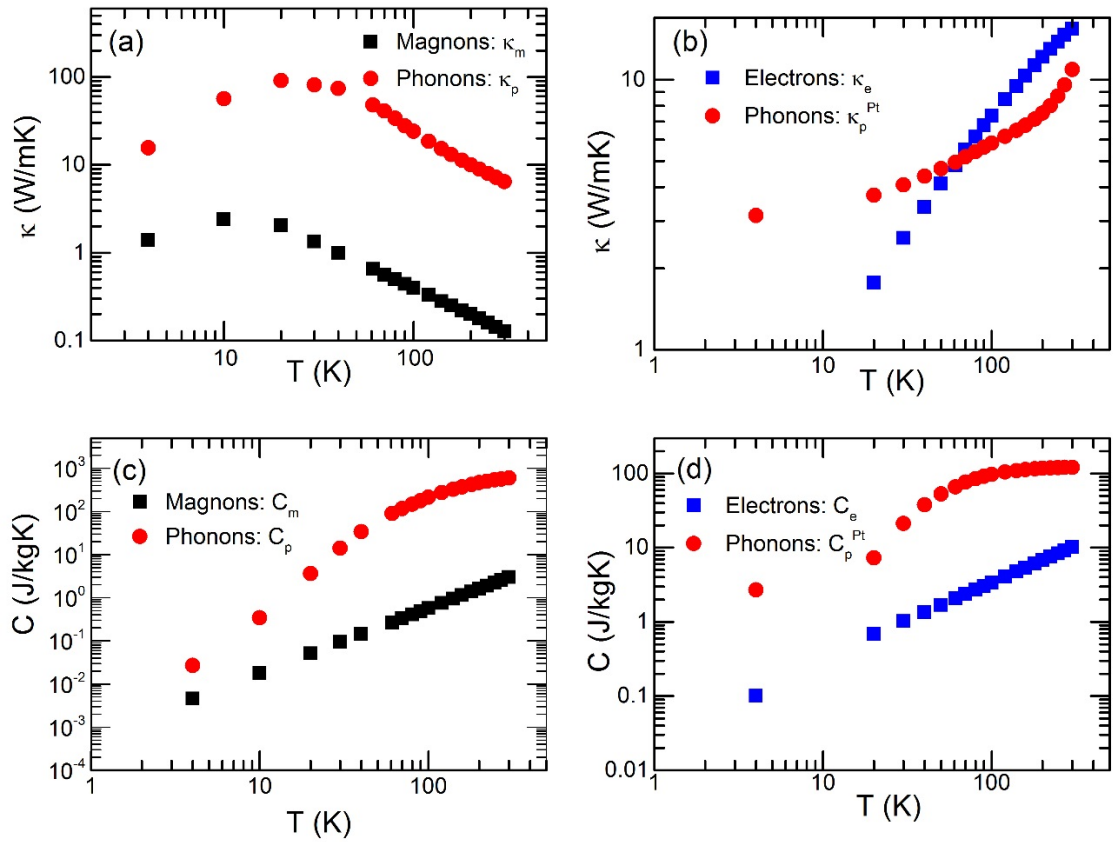

FIG. 10. (Color Online) (a) Magnon ($\kappa_m$) and phonon ($\kappa_p$) thermal conductivity values used for YIG in the FEM simulations. (b) Electron ($\kappa_e$) and phonon ($\kappa_p^{Pt}$) thermal conductivity values used for Pt. (c) Magnon ($C_m$) and phonon ($C_p$) heat capacities used for YIG. (d) Electron ($C_e$) and phonon ($C_p^{Pt}$) heat capacities used for Pt.

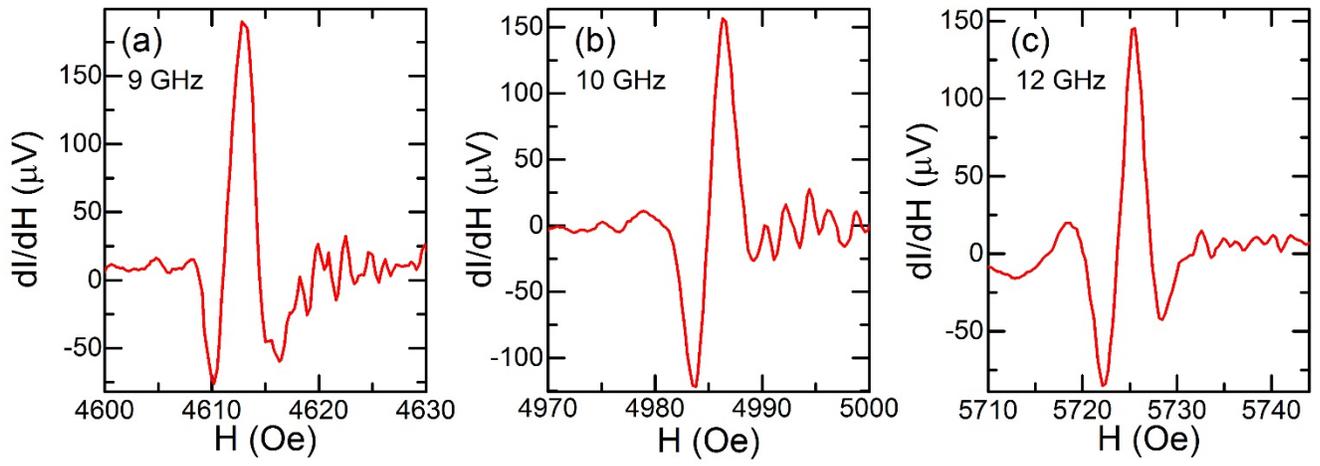

FIG. 11. (Color Online) FMR signal at 9 (a), 10 (b) and 12 (c) GHz.

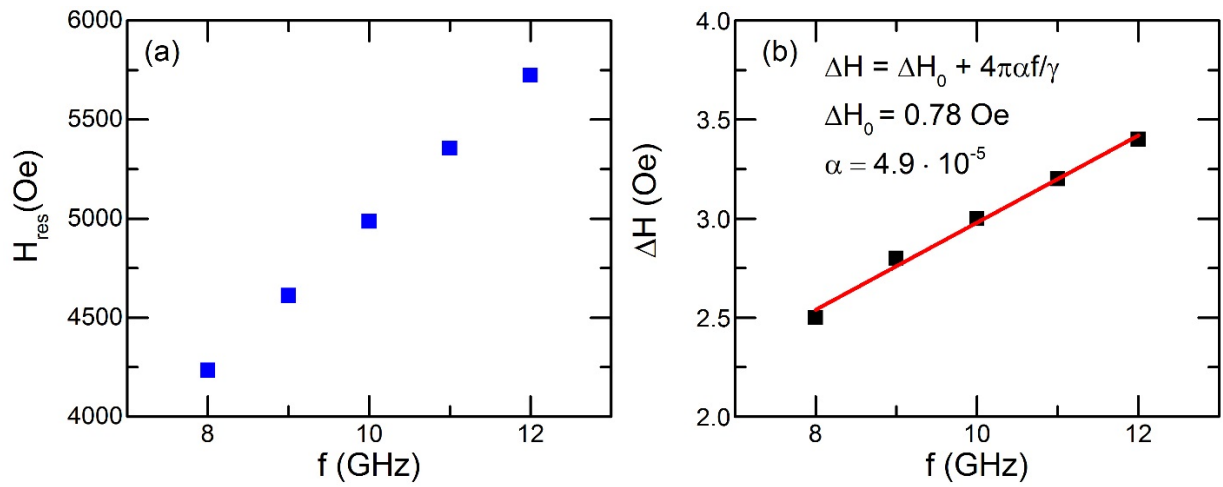

FIG. 12. (Color Online) (a) Frequency dependence of the resonant field $H_{res}$ from the stripline FMR measurements. (b) Frequency dependence of the FMR Linewidth (Lorentzian FWHM), the red line is the line of best fit.

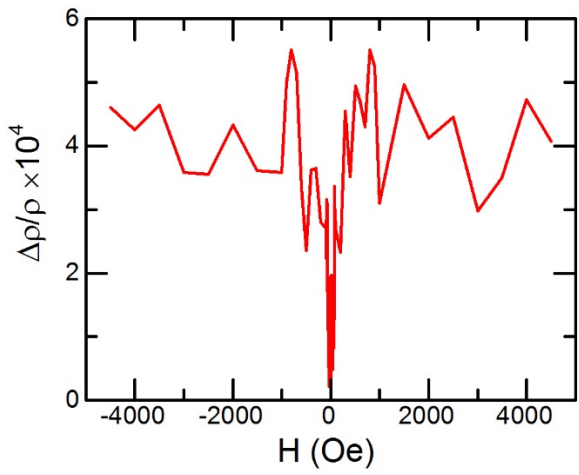

FIG. 13. (Color Online) Field dependence of the change in longitudinal resistivity ($\Delta\rho/\rho$) from the SMR measurements.

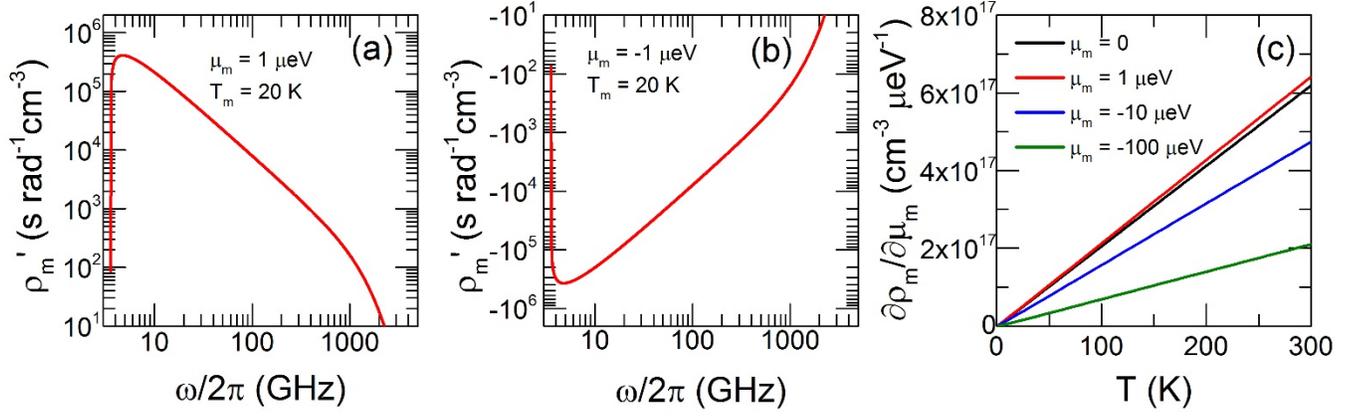

FIG. 14. (Color Online) Spectrally resolved non-equilibrium magnon density ($\rho_m'$) calculated from Eq. S11 at $T_m$ = 20 K and at $\mu_m$ = 1 μeV (a) and -1 μeV (b). $\partial \rho_m / \partial \mu_m$ versus temperature evaluated numerically at several values of $\mu_m$.

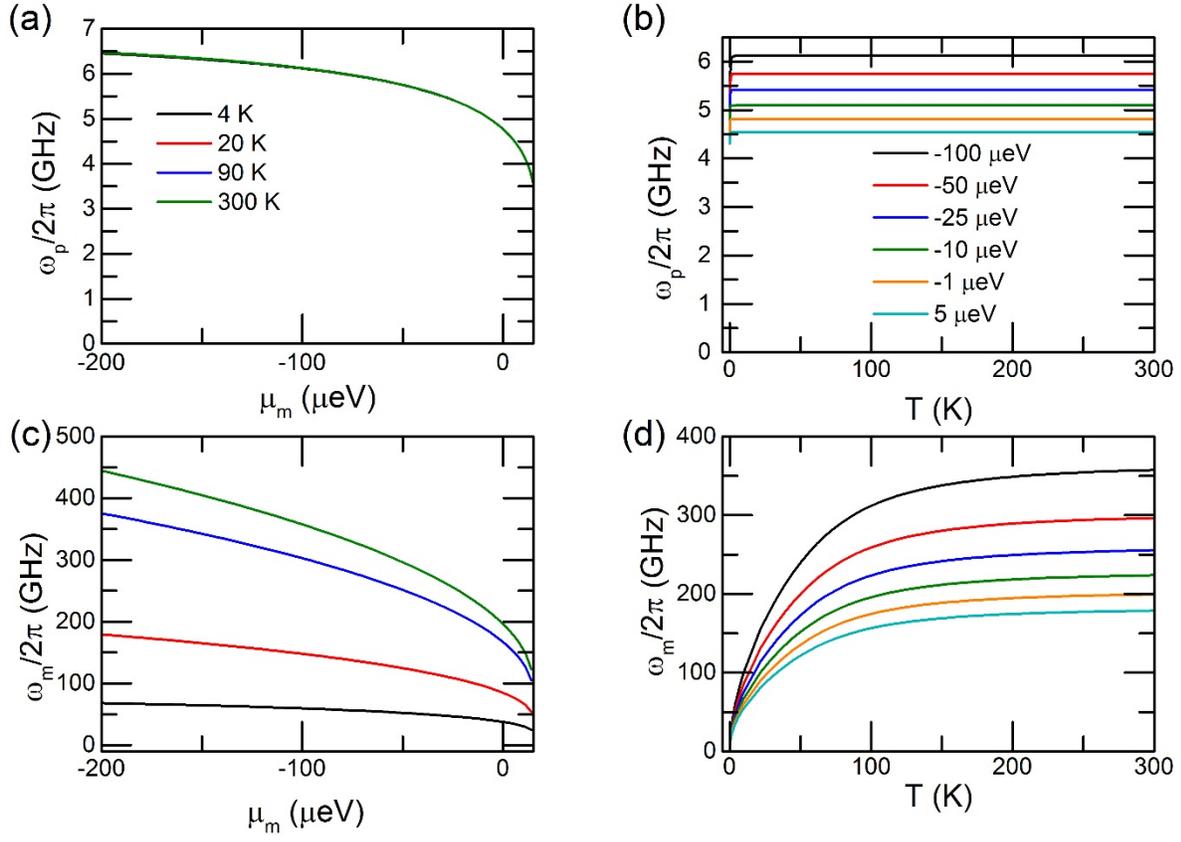

FIG. 15. (Color Online) Peak non-equilibrium magnon frequencies $\omega_p$ versus $\mu_m$ (a) and versus $T$ (b). Mean frequency of the non-equilibrium magnon distribution versus $\mu_m$ (c) and versus $T$ (d). (a,c) share a legend as do (b,d).

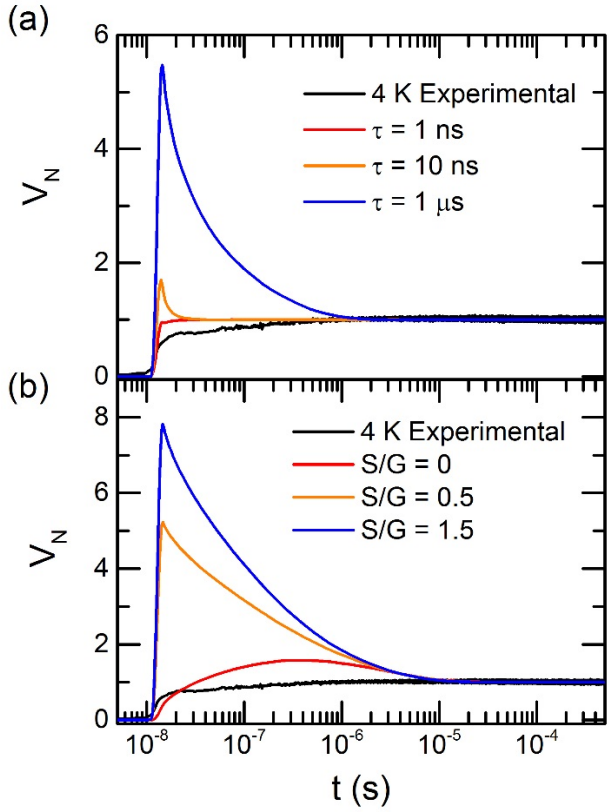

FIG. 16. (Color Online) Simulated $V_N$ at 4 K as a function of $\tau$ (a) and $S/G$ (b) using a linear temperature dependence on the Gilbert damping, $\alpha$. The shown simulations use $\alpha = 6.7 \times 10^{-7}$. In (a), the simulations use $\tau_{mp} = 200$ ps and $S/G = 1$. In (b), the simulations use $\tau = 10$ μs and $\tau_{mp} = 200$ ps.

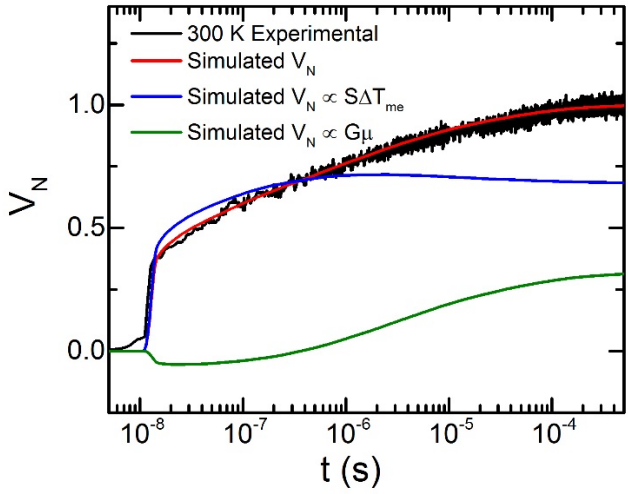

FIG. 17. (Color Online) Best fit at 300 K using $C_m = 14.2$ J/kgK. $\tau = 1\,ns$, $\tau_{mp} = 250\,\mu s$, $S/G = 9$

FIG. 1. (Color Online) (a) Schematic of the experimental setup. (b) $V_y$ measured vs. magnetic using a lock-in amplifier at 30K. (c) $V_y$ measured as a function of time and magnetic field using the oscilloscope at 30 K. Inset: Photodiode response to laser turning on at t = 10.8 ns and it's 10-90% rise time to be 2.2 ns and its 100% rise time to be 3 ns.

FIG. 2. (Color Online) (a) $V_{SSE}$ waveform vs time at various temperatures. (b) same data set as (a) normalized by $V_{SSE}(500\mu s)$ and plotted on a logarithmic time axis. Inset: $\tau_{SR}$ as a function of temperature the red horizontal line represents the 0-100% rise time of the laser.

FIG. 3. (Color Online) Zoom of $V_{SSE}$ versus time to show the negative $\partial V_{SSE}/\partial t$ exhibited over several hundred microseconds at 4 (a), 10 (b), 20 (c), and 30(d) K. The data in (a-d) are filtered using an adjacent averaging filter with a window width of 640 ns. (e) Drop in SSE signal ($\Delta V_{SSE}$) as a function of temperature as defined in the figure.

FIG. 4. (Color Online) Total magnon temperature excursion $\Delta T_m$ (a), z-component of the magnon temperature gradient (b), Laplacian of magnon temperature (c), and magnon phonon temperature difference (d) at 4 (black), 90 (red) and 300 K (blue). Data are taken from the FEM simulations at $r = 0$ and at $z = 0$.

FIG. 5. (Color Online) Contour maps of $\mu_m$ in YIG at $t = 50$ ns (a), and $t = 500$ μs (b) taken from the FEM simulations at the best fit. Note the different color scales and y axis ranges between (a) and (b). (c) z axis profiles of $\mu_m$ taken along the optical axis from the FEM simulations at various times.

FIG. 6. (Color Online) z-axis cuts of the bulk intrinsic SSE spin current (a), bulk non-equilibrium magnon spin current (b), and total bulk spin current (c) in YIG taken from the FEM simulations at the best fit at 90 K. (a-c) share a legend.

FIG. 7. (Color Online) Measured $V_N$ (black) at (a-b) 4 K and (c) 300 K. Modeled $V_N$ are plotted for different values of (a) magnon spin lifetime $\tau$ and (b) magnon-phonon thermalization time $\tau_{mp}$, in order to show the impact on the SSE waveform. (c) The simulated $V_N$ (red) is separated into the bulk $G\mu$ (green) and interface $S\Delta T_{me}$ (green) components.

FIG. 8. (Color Online) Temperature dependences of (a) magnon spin lifetime $\tau$, (b), magnon-phonon thermalization time $\tau_{mp}$, (c) magnon spin conductivity $\sigma_m$, and (d) pseudo-steady-state SSE voltage: measured $V_{SSE}$ (t = 500 μs) (black), modeled SSE due to $\mu_m$ (green), and $\Delta T_{me}$ (blue). Error bars are one standard error of the fits in each panel.

FIG. 9. (Color Online) (a) Spectrally resolved magnon relaxation times from Ref. [37]. (b) Spectrally resolved magnon spin conductivity calculated from a frequency dependent Eq. (8) and (a). (c) Cumulative magnon spin conductivity calculated from (b).

FIG. 10. (Color Online) (a) Magnon ($\kappa_m$) and phonon ($\kappa_p$) thermal conductivity values used for YIG in the FEM simulations. (b) Electron ($\kappa_e$) and phonon ($\kappa_p^{Pt}$) thermal conductivity values used for Pt. (c) Magnon ($C_m$) and phonon ($C_p$) heat capacities used for YIG. (d) Electron ($C_e$) and phonon ($C_p^{Pt}$) heat capacities used for Pt.

FIG. 11. (Color Online) FMR signal at 9 (a), 10 (b) and 12 (c) GHz.

FIG. 12. (Color Online) (a) Frequency dependence of the resonant field $H_{res}$ from the stripline FMR measurements. (b) Frequency dependence of the FMR Linewidth (Lorentzian FWHM), the red line is the line of best fit.

FIG. 13. (Color Online) Field dependence of the change in longitudinal resistivity ($\Delta\rho/\rho$) from the SMR measurements.

FIG. 14. (Color Online) Spectrally resolved non-equilibrium magnon density ($\rho_m'$) calculated from Eq. S11 at $T_m = 20$ K and at $\mu_m = 1$ μeV (a) and -1 μeV (b). $\partial\rho_m/\partial\mu_m$ versus temperature evaluated numerically at several values of $\mu_m$.

FIG. 15. (Color Online) Peak non-equilibrium magnon frequencies $\omega_p$ versus $\mu_m$ (a) and versus $T$ (b). Mean frequency of the non-equilibrium magnon distribution versus $\mu_m$ (c) and versus $T$ (d). (a,c) share a legend as do (b,d).

FIG. 16. (Color Online) Simulated $V_N$ at 4 K as a function of $\tau$ (a) and $S/G$ (b) using a linear temperature dependence on the Gilbert damping, $\alpha$. The shown simulations use $\alpha = 6.7 \times 10^{-7}$. In (a), the simulations use $\tau_{mp} = 200$ ps and $S/G = 1$. In (b), the simulations use $\tau = 10$ μs and $\tau_{mp} = 200$ ps.

FIG. 17. (Color Online) Best fit at 300 K using $C_m = 14.2$ J/kgK. $\tau = 1\,ns$, $\tau_{mp} = 250\,\mu s$, $S/G = 9$